# Coarse-grained Bootstrap of Quantum Many-body Systems

Minjae Cho[1,†], Colin Nancarrow[2,‡], Petar Tadić[3,5,6,§], Yuan Xin[4,5,¶], Zechuan Zheng[7,∥]

[1] *Kadanoff Center for Theoretical Physics & Enrico Fermi Institute, University of Chicago, Chicago, IL 60637, USA*

[2] *Department of Physics, Boston University, Boston, MA 02215, USA*

[3] *Maxwell Institute for Mathematical Sciences, Department of Mathematics, Heriot-Watt University, Edinburgh EH14, UK*

[4] *Department of Physics, Carnegie Mellon University, Pittsburgh, PA 15213, USA*

[5] *Department of Physics, Yale University, New Haven, CT 06511, USA*

[6] *Institute for Interdisciplinary and Multidisciplinary Studies, University of Montenegro, Podgorica, Montenegro*

[7] *Perimeter Institute for Theoretical Physics, Waterloo, ON N2L 2Y5, Canada*

We present a new computational framework combining coarse-graining techniques with bootstrap methods to study quantum many-body systems. The method efficiently computes rigorous upper and lower bounds on both zero- and finite-temperature expectation values of any local observables of infinite quantum spin chains. This is achieved by using tensor networks to coarse-grain bootstrap constraints, including positivity, translation invariance, equations of motion, and energy-entropy balance inequalities. Coarse-graining allows access to constraints from significantly larger subsystems than previously possible, yielding substantially tighter bounds compared to earlier methods.

---

[†] `cho7@uchicago.edu`
[‡] `nancarrow@bu.edu`
[§] `p.tadic@hw.ac.uk`
[¶] `yuanxin@andrew.cmu.edu`
[∥] `zechuan.zheng.phy@gmail.com`

# Contents

## 1 Introduction

Quantum many-body systems, while presenting intriguing physics, pose unique challenges in computing their physical observables. These systems contain infinitely many degrees of freedom that interact with one another nontrivially. Consequently, apart from a few special classes of integrable examples, systematic analytic approaches to generic strongly interacting quantum many-body systems are lacking. Even numerically, the cost of directly computing observables grows exponentially with system size, making such computations impractical. Thus, it has become important to develop alternative methods that are computationally efficient yet still capture the essential physics of quantum many-body systems.

One longstanding approach to this problem is coarse-graining. At the core of the renormalization group (RG) framework is the idea that only a subset of the ultraviolet-complete Hilbert space is relevant at low energies. One practical realization of this insight is the density matrix renormalization group (DMRG) [1], which in a broader context is known as the tensor network approach.[1] For one-dimensional lattice systems in particular, the matrix product state (MPS) ansatz for low-energy states has been shown, both analytically and numerically, to provide highly accurate approximations [3]. With a fixed bond dimension, the number of MPS parameters grows only linearly with system size, making the computation of observables much cheaper and faster.

---
[1]See e.g. [2] and references therein for a comprehensive review on the subject.

The concept of coarse-graining can be further extended to thermal equilibrium states at finite temperatures. In conjunction with the locality of the underlying Hamiltonian $H$, the thermal density matrix $\sim e^{-\beta H}$ at high temperatures (small $\beta$) lends itself to efficient tensor network approximations. For one-dimensional lattice systems, the matrix product operator (MPO) ansatz has been shown to provide an accurate approximation of the thermal density matrix at high temperatures [4]. This approach has also been generalized to projected entangled pair states (PEPS) in higher dimensional lattices [5] and convex combinations of MPS [6].

Another approach that has gained significant traction over the past two decades is the bootstrap method. Rather than focusing on wavefunctions, the bootstrap formulation imposes constraints that the observables must satisfy, such as the positivity of density matrices and the equations of motion (EOM) governing stationary states. While infinitely many independent constraints exist for an infinite system, the locality of the Hamiltonian allows for a systematic hierarchy among them, where only a relevant subset of constraints can be applied in practice. As these constraints must hold for the given system, the results obtained from the bootstrap method are rigorous, often yielding upper and lower bounds on observables of interest.

Since its initial success in quantum chemistry [7, 8], the bootstrap approach has found applications in areas including conformal bootstrap [9], lattice quantum field theories [10–16], classical and stochastic dynamical systems [17], quantum mechanical systems [18–28], and open quantum systems [29–32]. In many cases, the bootstrap approach can be formulated as a convex optimization problem, solvable using efficient softwares [33–36], offering a complementary tool to traditional methods such as exact diagonalization, tensor networks, and Monte Carlo simulations. For certain systems, rigorous theorems ensure that bootstrap results converge to the true physical values as more constraints are systematically incorporated.

Recent work [37] has shown that quantum systems on infinite lattices with finite-dimensional local state spaces provide another example where bootstrap results converge to physical values. In [37], the expectation values of local operators in both zero- and finite-temperature equilibrium states were studied using three main bootstrap constraints: positivity of density matrices, EOM, and energy-entropy balance (EEB) inequalities (which reduce to perturbative positivity at zero temperature). However, the size of the relevant convex optimization problem increases exponentially as the size of the sub-lattice on which bootstrap constraints are imposed grows. Thus, while the bootstrap method yields rigorous bounds, achieving highly precise results in practice remains computationally expensive.

A significant observation made in recent work [38] is that it is possible to combine the two approaches, coarse-graining and bootstrap, in a systematic way. Instead of imposing the full set of bootstrap constraints on a given sub-lattice, [38] proposed imposing only a relevant subset of constraints obtained by coarse-graining. In the case of one-dimensional lattice systems, they coarse-grained the reduced density matrices using MPS tensors and imposed locally translation-invariant (LTI) conditions on the coarse-grained reduced density matrices, alongside their positivity. These constraints, while coarse-grained, remain rigorous, and the computational cost is significantly reduced compared to the full bootstrap setup. This allows for constraints to be imposed on much larger sub-lattices, resulting in highly precise lower bounds on the ground state energies of spin-chain systems, surpassing previously reported results.

The goal of the current work is to extend the framework introduced in [38] by coarse-graining not only positivity constraints and LTI conditions but also EOM and EEB inequalities. While [38] provided only lower bounds on the ground state energy, our extension allows for both upper and lower bounds on zero- and finite-temperature expectation values of any local observables. We demonstrate that, by applying coarse-grained bootstrap constraints on larger sub-lattices, we can obtain significantly tighter bounds than previously achievable, showcasing the power of combining coarse-graining with the aforementioned bootstrap constraints of [37].

This paper is organized as follows. In section 2, we review two bootstrap approaches to quantum many-body systems: (1) constraining the equilibrium expectation values of local observables [37], and (2) constraining the matrix elements of ground-state reduced density matrices and their coarse-grained tensors [38]. In section 3, we describe how these two approaches can be combined to develop a coarse-grained equilibrium bootstrap framework, enabling the efficient bootstrap formulation of equilibrium expectation values for arbitrary local observables. Section 4 presents the corresponding numerical bootstrap results for the ground states of the transverse field Ising spin chain and the XXZ model. For the Ising model, we also present preliminary results at finite temperatures. Finally, we conclude with further discussions in section 5.

## 2 Review of Thermal Bootstrap and Coarse-Graining

In this section, we provide a brief review of two main ideas that will lead to the coarse-grained bootstrap setup. The first is the bootstrap for thermal equilibrium states as discussed in [37], whose essential observation was that the Kubo-Martin-Schwinger (KMS) conditions can be reformulated as convex conditions on the expectation values of local observables. The second is the coarse-graining of the bootstrap constraints for ground states introduced in [38] which provides a relaxed, but precise enough and rigorous bootstrap setup for quantum many-body systems. For the full details of this review section, we refer the readers to [37, 38].

### 2.1 Bootstrapping equilibrium states

KMS conditions are defining properties of a thermal equilibrium state. Given such a state at temperature $T = \frac{1}{\beta}$, it has been shown [39, 40][2] that KMS conditions are equivalent to the

---

[2]See also [41] and Appendix A.

following convex conditions on thermal expectation values $\langle\cdots\rangle_\beta$:

$$\text{Unit normalization: } \langle 1 \rangle_\beta = 1 \tag{2.1}$$

$$\text{Positivity of density matrix: } \langle \mathcal{O}^\dagger \mathcal{O} \rangle_\beta \geq 0, \quad \forall \mathcal{O} \tag{2.2}$$

$$\text{EOM: } \langle [H, \mathcal{O}] \rangle_\beta = 0, \quad \forall \mathcal{O} \tag{2.3}$$

$$\text{EEB inequalities: } \langle \mathcal{O}^\dagger \mathcal{O} \rangle_\beta \log\left(\frac{\langle \mathcal{O}^\dagger \mathcal{O} \rangle_\beta}{\langle \mathcal{O} \mathcal{O}^\dagger \rangle_\beta}\right) \leq \beta \langle \mathcal{O}^\dagger [H, \mathcal{O}] \rangle_\beta \quad \forall \mathcal{O}, \tag{2.4}$$

where $\mathcal{O}$ is any local observable.

At zero temperature ($\beta = \infty$), (2.4) is straightforward to derive:

$$\langle \Omega | \mathcal{O}^\dagger [H, \mathcal{O}] | \Omega \rangle = \sum_k \langle \Omega | \mathcal{O}^\dagger | k \rangle \langle k | [H, \mathcal{O}] | \Omega \rangle = \sum_k (E_k - E_0) |\langle \Omega | \mathcal{O}^\dagger | k \rangle|^2 \geq 0, \tag{2.5}$$

where $|\Omega\rangle$ is the ground state and $|k\rangle$ is a complete set of energy eigenstates with energy $E_k$. This leads to a straightforward SDP formulation of the ground states as follows:

**Ground state optimization**

$$\begin{aligned}
&\text{minimize/maximize } \langle \mathcal{O}_b \rangle \text{ subject to:} \\
&\text{Unit normalization: } \langle 1 \rangle = 1 \\
&\text{Positivity of density matrix: } \mathcal{A} \succeq 0, \quad \mathcal{A}_{ij} \equiv \langle \mathcal{O}_i^\dagger \mathcal{O}_j \rangle \\
&\text{EOM: } \langle [H, \mathcal{O}_i] \rangle = 0 \\
&\text{Perturbative positivity: } \mathcal{C} \succeq 0, \quad \mathcal{C}_{ij} \equiv \langle \mathcal{O}_i^\dagger [H, \mathcal{O}_j] \rangle,
\end{aligned} \tag{2.6}$$

where $\{\mathcal{O}_i\}$ provides a basis of local operators.

In contrast, at nonzero temperature, (2.4) presents an obstruction to a straightforward SDP formulation. Nonetheless, it can be formulated as a convex optimization as shown in [37, 42]. In practice, we would like to impose the bootstrap constraints on a finite-dimensional subalgebra of the algebra of all the local operators. Given such a finite-dimensional subalgebra of operators $\mathfrak{A}_f$, thermal expectation values $\langle\cdots\rangle_\beta$ must still satisfy all the conditions (2.2), (2.3), and (2.4) with the restriction $\mathcal{O} \in \mathfrak{A}_f$. Furthermore, by introducing the basis $\mathfrak{B}_f$ of $\mathfrak{A}_f$, EEB inequalities (2.4) can be equivalently re-written as

$$\beta \mathcal{C} + \mathcal{A}^{\frac{1}{2}} \log\left(\mathcal{A}^{-\frac{1}{2}} \mathcal{B} \mathcal{A}^{-\frac{1}{2}}\right) \mathcal{A}^{\frac{1}{2}} \succeq 0, \tag{2.7}$$

where $\mathcal{A}, \mathcal{B}$ and $\mathcal{C}$ matrices are defined as

$$\mathcal{A}_{ij} = \langle \mathcal{O}_i^\dagger \mathcal{O}_j \rangle_\beta, \ \mathcal{B}_{ij} = \langle \mathcal{O}_j \mathcal{O}_i^\dagger \rangle_\beta, \ \mathcal{C}_{ij} = \langle \mathcal{O}_i^\dagger [H, \mathcal{O}_j] \rangle_\beta, \ \mathcal{O}_{i,j} \in \mathfrak{B}_f. \tag{2.8}$$

The set of matrices $\mathcal{A}, \mathcal{B}$ and $\mathcal{C}$ subject to (2.7) form a convex cone, leading to the following convex optimization formulation of the thermal states:

**Thermal state optimization**

$$\begin{aligned}
&\text{minimize/maximize } \langle \mathcal{O}_b \rangle_\beta \text{ subject to:} \\
&\text{Unit normalization: } \langle 1 \rangle_\beta = 1, \\
&\text{Positivity of density matrix: } \mathcal{A} \succeq 0, \\
&\text{EOM: } \langle [H, \mathcal{O}_i] \rangle_\beta = 0, \\
&\text{EEB inequalities: } \beta \mathcal{C} + \mathcal{A}^{\frac{1}{2}} \log \left( \mathcal{A}^{-\frac{1}{2}} \mathcal{B} \mathcal{A}^{-\frac{1}{2}} \right) \mathcal{A}^{1/2} \succeq 0,
\end{aligned} \quad (2.9)$$

where matrices $\mathcal{A}, \mathcal{B}$ and $\mathcal{C}$ are given by (2.8). Practical implementation of (2.7) is achieved by the relevant self-concordant barrier function studied in [42], which has recently been employed on a numerical solver `qics` developed in [36].

In addition to the bootstrap constraints discussed above, there may be further constraints coming from symmetries, operator relations such as canonical commutations, and so on. In particular, for translation-invariant systems on the lattice, we expect the equilibrium states to be translation-invariant. Upon introducing $\tau_x$ which translates a local operator by one lattice spacing along the direction $x$, translation-invariance leads to the extra convex conditions on the equilibrium states:

$$\langle \tau_x(\mathcal{O}) \rangle_\beta = \langle \mathcal{O} \rangle_\beta, \quad \forall x, \; \forall \mathcal{O}. \quad (2.10)$$

## 2.2 Reduced density matrix bootstrap

Prior to the advent of the bootstrap formulation discussed in the previous section, there has been another bootstrap approach to the ground states of quantum mechanical systems consisting of many particles with pairwise interactions based on the positivity of reduced density matrices, as first introduced in the quantum chemistry literature [7, 8].

In the context of systems on the lattice, the reduced density matrix bootstrap can be formulated as follows [38]. Assume that the ground state is invariant under the translation on the lattice. The corresponding density matrix $\rho$ then allows for the construction of $M$-site reduced density matrix $\rho_M$ where all degrees of freedoms on the lattice sites except for $M$ consecutive sites are traced over. They should satisfy locally translation invariant (LTI) conditions

$$\rho_{M-1} = \text{Tr}_{\text{L}}(\rho_M) = \text{Tr}_{\text{R}}(\rho_M) \quad \text{for } M = 3, 4, \ldots, \quad (2.11)$$

where $\text{Tr}_{\text{L/R}}(\rho_M)$ means tracing over the degree of freedom on the left-/right-most site of $\rho_M$ respectively.

The Hamiltonian with the nearest-neighbor interactions[3] on the one-dimensional lattice can be expressed as $H = \sum_i h_{i,i+1}$ where the sum is over lattice sites and $h_{i,i+1}$ acts non-trivially only over the sites $i$ and $i + 1$. When the Hamiltonian is translation-invariant, the ground state energy density is given simply by $\text{Tr}(h\rho_2)$ with $h = h_{i,i+1}$ and $\rho_2$ being the 2-site reduced density matrix where all but $i$-th and $(i + 1)$-th sites are traced over. Then, the reduced density matrix bootstrap for the ground state at level $N$ is given by the following SDP problem:

---

[3] Generalizing to new few more site interactions is straightforward, and in this work we focus on nearest neighbor interaction for simplicity.

**Reduced density matrix bootstrap**

$$\begin{aligned}
&\text{minimize } \text{Tr}(h\rho_2) \text{ subject to:} \\
&\text{Normalization: } \text{Tr}(\rho_2) = 1 \\
&\text{Positivity of reduced density matrices: } \rho_M \succeq 0, \ M = 2, 3, ..., N \\
&\text{LTI conditions: } \rho_{M-1} = \text{Tr}_\text{L}(\rho_M) = \text{Tr}_\text{R}(\rho_M), \ M = 3, 4, ..., N,
\end{aligned} \quad (2.12)$$

where the variables are the elements of the density matrices $\rho_M$, $M = 2, 3, ..., N$. The resulting minimum $\text{Tr}(h\rho_2)$ provides a rigorous lower bound on the ground state energy density, which is non-decreasing as $N$ grows.

## 2.3 Coarse-grainined reduced density matrix bootstrap

The reduced density matrix bootstrap (2.12) is confronted with the problem that $\rho_N$ is a matrix of $\sim d^{2N}$ variables, where $d$ is the dimension of the local Hilbert space on each site, and for the spin-1/2 chain $d = 2$. In [38], this computational challenge was resolved by noting that one can systematically coarse-grain the bootstrap constraints such that the number of the bootstrap variables grow only linearly in $N$, as we briefly review below.[4]

Define a tensor

$$B_\mu^{IJ} = \ {}^I\!\!\underset{\mu}{\overset{\phantom{\mu}}{\bullet}}\!\!{}^J, \quad I, J = 1, ..., m, \ \mu = 1, 2, \cdots, d, \quad (2.13)$$

of dimension $m \times m \times d$, where $m$ is "bond dimension" denoting how much entanglement the coarse-grained system can have. The reduced density matrices $\rho_M$ with $M \geq 4$ are contracted with $B_\mu^{IJ}$ on sites $2, ..., M-1$ to produce the coarse-grained density matrix $\omega_M$ of dimension $(d \times m \times m \times d)^2$ as

$$\boxed{\omega_M}_{\nu_1\ I'\ J'\ \nu_M}^{\mu_1\ I\ J\ \mu_M} = \boxed{\rho_M}_{\nu_1\ I'\ \cdots\ J'\ \nu_N}^{\mu_1\ I\ \cdots\ J\ \mu_N}. \quad (2.14)$$

LTI conditions for $\omega$ matrices are given by

$$\boxed{\omega_4} = \boxed{\rho_3} \ , \quad \boxed{\omega_4} = \boxed{\rho_3} \ , \quad (2.15)$$

$$\boxed{\omega_{M+1}} = \boxed{\omega_M} \ , \quad \boxed{\omega_{M+1}} = \boxed{\omega_M} \ , \quad M \geq 4. \quad (2.16)$$

After fixing the tensor $B_\mu^{IJ}$ to be some constant tensor, the coarse-grained bootstrap is then formulated as an SDP problem

---

[4]We focus on the coarse-graining based on the uniform MPS (uMPS) in this work, but [38] discusses other types of coarse-graining schemes too.

**Coarse-grained reduced density matrix bootstrap**

> minimize $\text{Tr}(h\rho_2)$ subject to:
> Normalization: $\text{Tr}(\rho_2) = 1$
> Positivity of reduced density matrices: $\rho_M \succeq 0, \ M = 2, 3$
> Positivity of coarse-grained reduced density matrices: $\omega_M \succeq 0, \ M = 4, ..., N$
> LTI conditions: $\rho_2 = \text{Tr}_L(\rho_3) = \text{Tr}_R(\rho_3)$ and constraints in (2.15) and (2.16), (2.17)

where bootstrap variables are the matrix elements of $\rho_2, \rho_3$, and $\omega_M$ for $M = 4, ..., N$. Since the dimension of $\omega_M$ does not depend on $M$, the number of SDP variables grows only linearly in $N$. Furthermore, since we employed a relaxation of the original bootstrap constraints, the obtained minimum for $\text{Tr}(h\rho_2)$ still provides a rigorous lower bound on the ground state energy density.

The SDP formulation (2.17) produces very precise numerical results as demonstrated in [38]. However, it comes with a limitation that the only observable on which rigorous bounds can be obtained is the ground state energy density.

In the next section, we discuss how the computationally efficient coarse-graining idea of (2.17) can be combined with the original bootstrap problem of (2.6) and (2.9), providing rigorous bounds on any local observables both at zero and finite temperature.

## 3  Coarse-Grained Equilibrium Bootstrap

In this section, we examine the coarse-grained version of the optimization problem formulated in the previous section, specifically (2.6) for the ground state and (2.9) for the thermal state at an inverse temperature $\beta$.

### 3.1  Hierachy of constraints including equations of motions

In order to offer a slight change of perspective on the tensor network methods of [38], which will clarify our own approach, we review the LTI conditions (2.11) using graphical notation. All the states and operators mentioned hereafter should be considered as acting within the full Hilbert space, rather than being restricted to the Hilbert space truncated to $N$ sites. For example:

$$|\nu_1 \cdots \nu_N\rangle\langle\mu_1 \cdots \mu_N| \equiv \ldots \otimes \mathbb{I}_0 \otimes |\nu_1\rangle\langle\mu_1| \otimes \ldots \otimes |\nu_N\rangle\langle\mu_N| \otimes \mathbb{I}_{N+1} \otimes \ldots \quad (3.1)$$

i.e., it acts as an identity operator for sites outside the range $\{1, 2, ..., N\}$.

Given a density matrix $\rho$ and the operators defined above, the matrix element of the reduced density matrix for the $N$-site sub-system can be expressed as:

$$(\rho_N)^{\mu_1 \cdots \mu_N}_{\nu_1 \cdots \nu_N} = \text{Tr}\left(\rho |\nu_1 \cdots \nu_N\rangle\langle\mu_1 \cdots \mu_N|\right), \quad (3.2)$$

where the trace Tr is taken over the Hilbert space of the entire system. The LTI conditions

$$\delta^{\nu_0}_{\mu_0}(\rho_N)^{\mu_0 \cdots \mu_{N-1}}_{\nu_0 \cdots \nu_{N-1}} = \delta^{\nu_N}_{\mu_N}(\rho_N)^{\mu_1 \cdots \mu_N}_{\nu_1 \cdots \nu_N} = (\rho_{N-1})^{\mu_1 \cdots \mu_{N-1}}_{\nu_1 \cdots \nu_{N-1}} \quad (3.3)$$

can be expressed using graphical notation as:

$$\left[\begin{array}{c}\mu_1 \cdots \mu_{N-1} \\ \rho_N \\ \nu_1 \cdots \nu_{N-1}\end{array}\right] = \left[\begin{array}{c}\mu_1 \cdots \mu_{N-1} \\ \rho_N \\ \nu_1 \cdots \nu_{N-1}\end{array}\right] = \left[\begin{array}{c}\mu_1 \cdots \mu_{N-1} \\ \rho_{N-1} \\ \nu_1 \cdots \nu_{N-1}\end{array}\right] . \quad (3.4)$$

A brief reflection on (3.2) reveals the bridge between (2.12) and (2.9). On one hand, we have density matrices whose partial traces can be iteratively coarse-grained. On the other hand, we have the operators whose expectation values are constrained by EOM and EEB inequalities.

In order to write the EOM that constrain the images of the reduced density matrices under the action of the Hamiltonian, we need to define such an action. The Hamiltonian itself is local but acts on the global state. However, the commutator of two local operators $a$ and $b$ is nonvanishing only at the intersection of their supports. Consequently, it is useful to define a product $\odot$, called the *on-site product*, which retains only those terms that have common support. This is a sensible operation only insofar as it computes the intermediate minuend and subtrahend of a commutator. In terms of its action on our superoperators, we have:

$$\text{Tr}\left(\rho|\nu_1\cdots\nu_{N-1}\rangle\langle\mu_1\cdots\mu_{N-1}| \odot H\right) \equiv \sum_{i=0}^{N-1} \text{Tr}\left(\rho|\nu_1\cdots\nu_{N-1}\rangle\langle\mu_1\cdots\mu_{N-1}|h_{i,i+1}\right)$$
$$= (H \odot \rho_{N-1})_{\nu_1\cdots\nu_{N-1}}^{\mu_1\cdots\mu_{N-1}}, \quad (3.5)$$

where $h_{i,i+1}$ denotes the two-site interaction on the $i$-th site. Due to locality, the vacuum-state equation of motion for $|\nu_1\cdots\nu_{N-1}\rangle\langle\mu_1\cdots\mu_{N-1}|$ can involve only the on-site product between $H$ and $\rho_{N-1}$:

$$0 = \text{Tr}\left(\rho\left[H, |\nu_1\cdots\nu_{N-1}\rangle\langle\mu_1\cdots\mu_{N-1}|\right]\right) = \rho_{N-1} \odot H - H \odot \rho_{N-1}, \quad (3.6)$$

where the subtrahend is given in diagrammatic notation by:

$$\left[\begin{array}{c}\mu_1 \cdots \mu_{N-1} \\ H \odot \rho_{N-1} \\ \nu_1 \cdots \nu_{N-1}\end{array}\right] = \left[\begin{array}{c}\mu_1 \cdots \mu_{N-1} \\ H_{\text{bulk}} \\ \rho_{N-1} \\ \nu_1 \cdots \nu_{N-1}\end{array}\right] + \left[\begin{array}{c}\mu_1 \\ h \\ \rho_N \\ \nu_1 \cdots \nu_{N-1}\end{array}\right] + \left[\begin{array}{c}\mu_{N-1} \\ h \\ \rho_N \\ \nu_1 \cdots \nu_{N-1}\end{array}\right] . \quad (3.7)$$

The first term is shorthand for the sum:

$$\begin{array}{c} \mu_1 \quad\quad \mu_{N-1} \\ \boxed{H_{\text{bulk}}} \\ \boxed{\rho_{N-1}} \\ \nu_1 \quad\quad \nu_{N-1} \end{array} = \sum_{i=1}^{N-2} \begin{array}{c} \mu_1 \quad \mu_i\,\mu_{i+1} \quad \mu_{N-1} \\ \boxed{h} \\ \boxed{\rho_{N-1}} \\ \nu_1 \quad\quad\quad \nu_{N-1} \end{array}. \tag{3.8}$$

Identifying $\rho_{N-1} \odot H$ as $(H \odot \rho_{N-1})^\dagger$, we can now interpret the hermiticity condition

$$(\rho_{N-1} \odot H) - (\rho_{N-1} \odot H)^\dagger = 0 \tag{3.9}$$

as a result of the quantum equation of motion through (3.7). That is, partial traces on $N$-site operators are related to $(N-1)$-site operators. Substituting (3.7) into (3.9) for $N$ ranging from 1 to $n$, we obtain a hierarchy of constraints that may be iteratively coarse-grained.

### 3.2 Coarse-graining the equations of motions

Following [38], we utilize uMPS as our coarse-graining maps. The uMPS was chosen by minimizing the ground state energy using the variational-uniform-MPS (VUMPS) algorithm in [43]. The $d^N$ different $|\mu_1 \ldots \mu_N\rangle$ states, whose outer products previously formed the superoperators, are now replaced by the $m^2$ coarse-grained states:

$$|\psi_N^{IJ}\rangle \equiv \sum_{\{\mu_i\}} B_{\mu_1}^{IK_1} B_{\mu_2}^{K_1 K_2} \cdots B_{\mu_N}^{K_{N-1} J} |\mu_1 \mu_2 \cdots \mu_N\rangle = \begin{array}{c} I \;\text{---}\;\bullet\;\text{---}\;\bullet\;\cdots\;\bullet\;\text{---}\; J \\ |\quad |\quad\quad | \\ \mu_1 \;\; \mu_2 \quad\quad \mu_N \end{array}, \tag{3.10}$$

where $B_\mu^{IJ}$ is the coarse-graining tensor with bond dimension $m$. The LTI conditions require taking the partial trace of peripheral tensor legs, so we will also employ coarse-graining maps that leave such legs untouched. For example:

$$|\mu_1 \psi_{N-2}^{IJ} \mu_N\rangle \equiv \sum_{\mu_2 \cdots \mu_{N-1}} B_{\mu_2}^{IK_1} B_{\mu_3}^{K_1 K_2} \cdots B_{\mu_{N-1}}^{K_{N-3} J} |\mu_1 \mu_2 \cdots \mu_N\rangle. \tag{3.11}$$

Coarse-grained vacuum expectation values may now be defined via:

$$(\omega_{L,N})_{\nu_1 \nu_2 I' J' \nu_N}^{\mu_1 \mu_2 I J \mu_N} \equiv \text{Tr}\left(\rho |\nu_1 \nu_2 \psi_{N-3}^{I'J'} \nu_N\rangle \langle \mu_1 \mu_2 \psi_{N-3}^{IJ} \mu_N|\right) = \begin{array}{c} \mu_1\,\mu_2\; I \quad\quad J\; \mu_N \\ \boxed{\rho_N} \\ \nu_1\,\nu_2\; I' \quad\quad J'\; \nu_N \end{array},$$

$$(\omega_{R,N})_{\nu_1 I' J' \nu_{N-1} \nu_N}^{\mu_1 I J \mu_{N-1} \mu_N} \equiv \text{Tr}\left(\rho |\nu_1 \psi_{N-3}^{I'J'} \nu_{N-1} \nu_N\rangle \langle \mu_1 \psi_{N-3}^{IJ} \mu_{N-1} \mu_N|\right) = \begin{array}{c} \mu_1\; I \quad\quad J\,\mu_{N-1}\mu_N \\ \boxed{\rho_N} \\ \nu_1\; I' \quad\quad J'\,\nu_{N-1}\nu_N \end{array},$$

(3.12)

$$(\chi_N)_{\nu_1 I' J' \nu_N}^{\mu_1 I J \mu_N} \equiv \sum_{i=1}^{N-1} \mathrm{Tr}\left(\rho |\nu_1 \psi_{N-2}^{I'J'} \nu_N\rangle \langle \mu_1 \psi_{N-2}^{IJ} \mu_N | h_{i,i+1}|\right) = \begin{array}{c} \includegraphics \end{array}. \tag{3.13}$$

The key point expressed in (3.12) and (3.13) is that, as far as the optimization problem is concerned, the only parameters that appear are the *images* of the coarse-graining maps. For example, the tensor $(\chi_N)_{\nu_1 I' J' \nu_N}^{\mu_1 I J \mu_N}$ collects only $d^4 \times m^4$ parameters as opposed to $d^{2N}$. This is also the reason we must define both $\chi$ and $\omega$: after coarse-graining, there is no direct way to express $\chi$ in terms of $\omega$ using $H$. This must be done step-by-step, as we shall see. We have avoided coarse-graining not just two but three boundary sites in the $\omega$ tensors so that we may continue to act with the two-site operator $h$ as in (3.7). The LTI conditions for these tensors are:

$$\omega_{L,N} = \omega_{R,N} = \omega_{C,N-1} \quad, \tag{3.14}$$

$$\omega_{L,N} = \omega_{L,N-1} \quad, \tag{3.15}$$

$$\omega_{R,N} = \omega_{R,N-1} \quad, \tag{3.16}$$

where we have defined for convenience the object $\omega_{C,N}$ by

$$\omega_{C,N} \equiv \omega_{L,N} = \omega_{R,N}. \tag{3.17}$$

The $\chi_N$ tensors contain information from the Hamiltonian. The LTI conditions they satisfy

are:

$$\chi_N \Big|_{\nu_1\ I'\ J'}^{\mu_1\ I\ J} = \chi_{N-1} \Big|_{\nu_1\ I'\ J'}^{\mu_1\ I\ J} \bullet_{J'}^{J} + \omega_{R,N} \Big|_{\nu_1\ I'\ J'}^{\mu_1\ I\ \boxed{h}^J} \quad , \tag{3.18}$$

$$\chi_N \Big|_{I'\ J'\ \nu_N}^{I\ J\ \mu_N} = \chi_{N-1} \Big|_{I'\ J'\ \nu_N}^{I\ J\ \mu_N} \bullet_{I'}^{I} + \omega_{L,N} \Big|_{I'\ J'\ \nu_N}^{\boxed{h}^I\ J\ \mu_N} \quad .$$

It is not difficult to see where these arise from: every two-site interaction in the Hamiltonian leaves the rightmost (leftmost) leg of $\rho_N$ unchanged, except for the final (initial) one. Therefore, the first term on the right-hand side of (3.18) is merely a consequence of $\rho_N$'s LTI condition, while the second term accounts for the residual boundary coupling. Note that $\chi_N$ is not Hermitian. We are now in a position to formulate the coarse-grained EOM:

$$0 = \operatorname{Tr}\left(\rho\big[|\nu_1\psi_{N-3}^{I'J'}\nu_{N-1}\rangle\langle\mu_1\psi_{N-3}^{IJ}\mu_{N-1}|, H\big]\right) = H \odot \omega_{C,N-1} - (H \odot \omega_{C,N-1})^\dagger \tag{3.19}$$

where

$$H \odot \omega_{C,N-1} \Big|_{\nu_1\ I'\ J'\ \nu_{N-1}}^{\mu_1\ I\ J\ \mu_{N-1}} = \chi_{N-1} \Big|_{\nu_1\ I'\ J'\ \nu_{N-1}}^{\mu_1\ I\ J\ \mu_{N-1}} + \omega_{L,N} \Big|_{\nu_1\ I'\ J'\nu_{N-1}}^{\boxed{h}\ I\ J\mu_{N-1}} + \omega_{R,N} \Big|_{\nu_1\ I'\ J'\nu_{N-1}}^{\mu_1\ I\ J\ \boxed{h}} . \tag{3.20}$$

At this point, we have everything required to implement the coarse-grained version of (3.7), along with the additional stationary state constraints. To target order parameters and correlation functions, rather than just the ground state energy density, we will proceed to formulate a coarse-grained version of EEB inequalities in the following sections.

### 3.3 Coarse-graining the perturbative positivity and thermal matrices

In this part, we discuss the coarse-graining of the EEB inequalities Eq. (2.7). Taking the limit $\beta \to \infty$, the EEB inequalities reduce to the following condition for the ground state:

$$\langle \mathcal{O}^\dagger [H, \mathcal{O}] \rangle \geq 0, \tag{3.21}$$

which implies that the matrix with elements $\mathcal{C}_{ij} = \langle \mathcal{O}_i^\dagger [H, \mathcal{O}_j] \rangle$ must be positive semidefinite. The construction and coarse-graining of this matrix for our basis of superoperators $|\nu_1 \cdots \nu_N\rangle\langle\mu_1 \cdots \mu_N|$ within the ground state is the key task we must complete to leverage the EEB inequalities. The term involving the logarithm can then be computed using the methods of [37].

We have as our positivity matrix

$$\begin{aligned}
\mathcal{C}^{(\{\lambda_i\},\{\mu_i\})}_{(\{\sigma_i\},\{\nu_i\})} &= \mathrm{Tr}\left(\rho|\nu_1\cdots\nu_{N-1}\rangle\langle\lambda_1\cdots\lambda_{N-1}|\left[H,|\sigma_1\cdots\sigma_{N-1}\rangle\langle\mu_1\cdots\mu_{N-1}|\right]\right) \\
&= \mathrm{Tr}\left(\rho|\nu_1\cdots\nu_{N-1}\rangle\langle\lambda_1\cdots\lambda_{N-1}|H \odot |\sigma_1\cdots\sigma_{N-1}\rangle\langle\mu_1\cdots\mu_{N-1}|\right) \\
&\quad - \mathrm{Tr}\left(\rho|\nu_1\cdots\nu_{N-1}\rangle\langle\lambda_1\cdots\lambda_{N-1}|\sigma_1\cdots\sigma_{N-1}\rangle\langle\mu_1\cdots\mu_{N-1}| \odot H\right)
\end{aligned} \qquad (3.22)$$

where we regard the collection of indices $(\{\lambda_i\},\{\mu_i\})$ and $(\{\sigma_i\},\{\nu_i\})$ as two super-indices for the purpose of stating positive semidefiniteness. In diagrammatic notation, we have for the first on-site product

$$\begin{aligned}
&\mathrm{Tr}\left(\rho|\nu_1\cdots\nu_{N-1}\rangle\langle\lambda_1\cdots\lambda_{N-1}|H \odot |\sigma_1\cdots\sigma_{N-1}\rangle\langle\mu_1\cdots\mu_{N-1}|\right) \\
&= \sum_{i=1}^{N-2} \delta^{\lambda_1}_{\sigma_1}\cdots\delta^{\lambda'_i}_{\sigma_i}\delta^{\lambda'_{i+1}}_{\sigma_{i+1}}\cdots\delta^{\lambda_{N-1}}_{\sigma_{N-1}} H^{\lambda_i\lambda_{i+1}}_{\lambda'_i\lambda'_{i+1}} (\rho_{N-1})^{\mu_1\cdots\mu_{N-1}}_{\nu_1\cdots\nu_{N-1}} \\
&\quad + H^{\nu_0\lambda_1}_{\mu_0\sigma_1}\delta^{\lambda_2}_{\sigma_2}\cdots\delta^{\lambda_{N-1}}_{\sigma_{N-1}} (\rho_N)^{\mu_0\mu_1\cdots\mu_{N-1}}_{\nu_0\nu_1\cdots\nu_{N-1}} + H^{\lambda_{N-1}\nu_N}_{\sigma_{N-1}\mu_N}\delta^{\lambda_1}_{\sigma_1}\cdots\delta^{\lambda_{N-2}}_{\sigma_{N-2}} (\rho_N)^{\mu_1\cdots\mu_{N-1}\mu_N}_{\nu_1\cdots\nu_{N-1}\nu_N}
\end{aligned}$$

$$= \begin{array}{c}\lambda_1 \cdots \lambda_{N-1} \\ \boxed{H} \\ \sigma_1 \cdots \sigma_{N-1}\end{array} \quad \begin{array}{c}\mu_1 \cdots \mu_{N-1} \\ \boxed{\rho_{N-1}} \\ \nu_1 \cdots \nu_{N-1}\end{array} \qquad (3.23)$$

$$+ \begin{array}{c}\lambda_2\ \lambda_{N-1} \\ \cdots \\ \sigma_2\ \sigma_{N-1}\end{array}\boxed{\begin{array}{c}\sigma_1 \\ h \\ \lambda_1\end{array}}\begin{array}{c}\mu_1 \cdots \mu_{N-1} \\ \boxed{\rho_N} \\ \nu_1 \cdots \nu_{N-1}\end{array} + \begin{array}{c}\lambda_1\ \lambda_{N-2} \\ \cdots \\ \sigma_1\ \sigma_{N-2}\end{array}\begin{array}{c}\mu_1 \cdots \mu_{N-1} \\ \boxed{\rho_N} \\ \nu_1 \cdots \nu_{N-1}\end{array}\boxed{\begin{array}{c}\sigma_{N-1} \\ h \\ \lambda_{N-1}\end{array}}.$$

The second term in (3.22) is easy to compute: the inner product between $\{\lambda_i\}$ and $\{\sigma_i\}$ becomes a $\delta$ symbol and what is leftover is just the on-site product:

$$\begin{aligned}
&\mathrm{Tr}\left(\rho|\nu_1\cdots\nu_{N-1}\rangle\langle\lambda_1\cdots\lambda_{N-1}|\sigma_1\cdots\sigma_{N-1}\rangle\langle\mu_1\cdots\mu_{N-1}| \odot H\right) \\
&= \delta^{\lambda_1}_{\sigma_1}\cdots\delta^{\lambda_{N-1}}_{\sigma_{N-1}} (H \odot \rho_{N-1})^{\mu_1\cdots\mu_{N-1}}_{\nu_1\cdots\nu_{N-1}}.
\end{aligned} \qquad (3.24)$$

The coarse-graining can now be performed exactly as in the prequel. Contracting an appropriate MPS of the form (3.10) on the tensor legs indexed with subscripts from $\{2,\ldots N-2\}$, we obtain

$$
\mathcal{C}^{(\lambda_1,K,L,\lambda_{N-1},\mu_1,I,J,\mu_{N-1})}_{(\sigma_1,K',L',\sigma_{N-1},\nu_1,I',J',\nu_{N-1})}
$$

$$
= \mathrm{Tr}\left(\rho|\nu_1\psi^{I'J'}_{N-3}\nu_{N-1}\rangle\langle\lambda_1\psi^{KL}_{N-3}\lambda_{N-1}|\left[H,|\sigma_1\psi^{K'L'}_{N-3}\sigma_{N-1}\rangle\langle\mu_1\psi^{IJ}_{N-3}\mu_{N-1}|\right]\right)
$$

$$
= \mathrm{Tr}\left(\rho|\nu_1\psi^{I'J'}_{N-3}\nu_{N-1}\rangle\langle\lambda_1\psi^{KL}_{N-3}\lambda_{N-1}|H\odot|\sigma_1\psi^{K'L'}_{N-3}\sigma_{N-1}\rangle\langle\mu_1\psi^{IJ}_{N-3}\mu_{N-1}|\right)
$$

$$
- \mathrm{Tr}\left(\rho|\nu_1\psi^{I'J'}_{N-3}\nu_{N-1}\rangle\langle\lambda_1\psi^{KL}_{N-3}\lambda_{N-1}|\sigma_1\psi^{K'L'}_{N-3}\sigma_{N-1}\rangle\langle\mu_1\psi^{IJ}_{N-3}\mu_{N-1}|\odot H\right)
$$

(3.25)

[Tensor network diagrams showing the decomposition: the first line contains a transfer matrix block with indices $K,L,K',L'$ multiplied by the sum of two diagrams (one with $h$ and $\omega_{L,N}$, one with $\omega_{R,N}$ and $h$), plus a second line with $H_{\text{bulk}}$ and $\omega_{C,N-1}$ terms minus a transfer matrix times $H\odot\omega_{C,N-1}$.]

where the transfer matrix

$$
\langle\psi^{KL}_{N-3}|\psi^{K'L'}_{N-3}\rangle = \quad \text{[diagram]} \quad . \tag{3.26}
$$

can be computed for a given $B$ tensor with the typical tensor network contraction algorithms.

For the thermal state at inverse temperature $\beta$, we further require the coarse-graining of the following two matrices, which corresponds to the $\langle\mathcal{O}^\dagger_i\mathcal{O}_j\rangle$ and $\langle\mathcal{O}_j\mathcal{O}^\dagger_i\rangle$ discussed in section 2:

$$
\begin{aligned}
\mathcal{A}^{(\{\lambda_i\},\{\mu_i\})}_{(\{\sigma_i\},\{\nu_i\})} &= \mathrm{Tr}\left(\rho|\nu_1\cdots\nu_{N-1}\rangle\langle\lambda_1\cdots\lambda_{N-1}|\sigma_1\cdots\sigma_{N-1}\rangle\langle\mu_1\cdots\mu_{N-1}|\right) \\
&= \delta^{\lambda_1}_{\sigma_1}\ldots\delta^{\lambda_{N-1}}_{\sigma_{N-1}}(\rho_{N-1})^{\mu_1\cdots\mu_{N-1}}_{\nu_1\cdots\nu_{N-1}}
\end{aligned}
\tag{3.27}
$$

$$
\begin{aligned}
\mathcal{B}^{(\{\lambda_i\},\{\mu_i\})}_{(\{\sigma_i\},\{\nu_i\})} &= \mathrm{Tr}\left(\rho|\sigma_1\cdots\sigma_{N-1}\rangle\langle\mu_1\cdots\mu_{N-1}|\nu_1\cdots\nu_{N-1}\rangle\langle\lambda_1\cdots\lambda_{N-1}|\right) \\
&= \delta^{\mu_1}_{\nu_1}\ldots\delta^{\mu_{N-1}}_{\nu_{N-1}}(\rho_{N-1})^{\lambda_1\cdots\lambda_{N-1}}_{\sigma_1\cdots\sigma_{N-1}}
\end{aligned}
\tag{3.28}
$$

The coarse-graining procedure goes in parallel with the matrix $\mathcal{C}$ from (3.22) to (3.25):

$$
\mathcal{A}^{(\lambda_1,K,L,\lambda_{N-1},\mu_1,I,J,\mu_{N-1})}_{(\sigma_1,K',L',\sigma_{N-1},\nu_1,I',J',\nu_{N-1})}
$$

$$
= \mathrm{Tr}\left(\rho|\nu_1\psi^{I'J'}_{N-3}\nu_{N-1}\rangle\langle\lambda_1\psi^{KL}_{N-3}\lambda_{N-1}|\sigma_1\psi^{K'L'}_{N-3}\sigma_{N-1}\rangle\langle\mu_1\psi^{IJ}_{N-3}\mu_{N-1}|\right)
\tag{3.29}
$$

$$
= \delta^{\lambda_1}_{\sigma_1}\delta^{\lambda_{N-1}}_{\sigma_{N-1}}\langle\psi^{KL}_{N-3}|\psi^{K'L'}_{N-3}\rangle(\omega_{C,N-1})^{\mu_1 IJ\mu_{N-1}}_{\nu_1 I'J'\nu_{N-1}}
$$

$$\mathcal{B}^{(\lambda_1,K,L,\lambda_{N-1},\mu_1,I,J,\mu_{N-1})}_{(\sigma_1,K',L',\sigma_{N-1},\nu_1,I',J',\nu_{N-1})}$$

$$=\mathrm{Tr}\left(\rho\big|\sigma_1\psi^{K'L'}_{N-3}\sigma_{N-1}\rangle\langle\mu_1\psi^{IJ}_{N-3}\mu_{N-1}|\nu_1\psi^{I'J'}_{N-3}\nu_{N-1}\rangle\langle\lambda_1\psi^{KL}_{N-3}\lambda_{N-1}\big|\right) \quad (3.30)$$

$$=\delta^{\mu_1}_{\nu_1}\delta^{\mu_{N-1}}_{\nu_{N-1}}\langle\psi^{IJ}_{N-3}|\psi^{I'J'}_{N-3}\rangle\,(\omega_{C,N-1})^{\lambda_1 KL\lambda_{N-1}}_{\sigma_1 K'L'\sigma_{N-1}}$$

### 3.4 Hierarchy of coarse-grained optimization

First, we present the optimization problem with no coarse-graining, given by (2.6) for the ground state and (2.9) for the thermal state, using the new notation and organization of the operators:

$$\begin{aligned}
\text{minimize:} \quad & \text{any convex objective functions over the variables under investigation,} \\
\text{over:} \quad & (\rho_r)^{\mu_1\cdots\mu_r}_{\nu_1\cdots\nu_r},\ r=1,2,\cdots,N, \\
\text{with constraints:} \quad & \rho_r \succeq 0, \\
& \mathrm{Tr}_1\rho_r = \mathrm{Tr}_r\rho_r = \rho_{r-1}, \\
& \mathrm{Tr}\rho_1 = 1, \\
& \text{LTI condition: (3.4),} \\
& \text{EOM: (3.6),} \\
& \text{EEB inequalities:} \quad \mathcal{C} \succeq -\tfrac{1}{\beta}\mathcal{A}^{\frac{1}{2}}\log\left(\mathcal{A}^{-\frac{1}{2}}\mathcal{B}\mathcal{A}^{-\frac{1}{2}}\right)\mathcal{A}^{1/2}, \\
& \text{for ground state, i.e. } \beta\to\infty,\quad \mathcal{C}\succeq 0,
\end{aligned} \quad (3.31)$$

where matrices $\mathcal{A}$, $\mathcal{B}$ and $\mathcal{C}$ are given by (3.27), (3.28) and (3.22), respectively.

Now, we summarize the coarse-grained optimization problem for the ground and thermal state at inverse temperature $\beta$ as follows:

$$\begin{aligned}
\text{minimize:} \quad & \text{any convex objective functions over the variables under investigation,} \\
\text{over:} \quad & (\rho_r)^{\mu_1\cdots\mu_r}_{\nu_1\cdots\nu_r},\ r=1,2,3,4, \\
& (\omega_{L,r})^{\mu_1\mu_2 IJ\mu_r}_{\nu_1\nu_2 I'J'\nu_r},\ (\omega_{R,r})^{\mu_1 IJ\mu_{r-1}\mu_r}_{\nu_1 I'J'\nu_{r-1}\nu_r},\ (\omega_{C,r})^{\mu_1 IJ\mu_r}_{\nu_1 I'J'\nu_r},\ r=5,6,\cdots,N, \\
& (\chi_r)^{\mu_1 IJ\mu_r}_{\nu_1 I'J'\nu_r},\ r=5,6,\cdots,N, \\
\text{with constraints:} \quad & \rho_r \succeq 0,\ \text{for } r\leq 4, \\
& \omega_{L,r}\succeq 0,\quad \omega_{R,r}\succeq 0,\quad \text{for } r\geq 5, \\
& \mathrm{Tr}_1\rho_r = \mathrm{Tr}_r\rho_r = \rho_{r-1},\quad \text{for } r\leq 4, \\
& \mathrm{Tr}\rho_1 = 1, \\
& \text{coarse-grained LTI condition: (3.14), (3.15), (3.16) for } r\geq 5, \\
& \text{EOM \& coarse-grained EOM: (3.6) for } r\leq 4 \text{ and (3.19) for } r\geq 5, \\
& \text{EEB inequalities:} \quad \mathcal{C}\succeq -\tfrac{1}{\beta}\mathcal{A}^{\frac{1}{2}}\log\left(\mathcal{A}^{-\frac{1}{2}}\mathcal{B}\mathcal{A}^{-\frac{1}{2}}\right)\mathcal{A}^{1/2}, \\
& \text{for ground state, i.e. } \beta\to\infty,\quad \mathcal{C}\succeq 0.
\end{aligned} \quad (3.32)$$

We note that the EEB inequalities here apply for $r\leq 4$ and $r\geq 5$. For each $r\leq 4$, we use the non-coarse-grained versions of $\mathcal{A}$, $\mathcal{B}$ and $\mathcal{C}$, as defined in (3.27), (3.28) and (3.22), respectively. For each $5\leq r\leq N-1$, we employ the coarse-grained versions as given in (3.29), (3.30) and (3.25).

We make an important remark here that, even though MPS is known to well-approximate the ground states of local Hamiltonian systems [3], it is in general not expected to provide a good approximation to thermal states. For this reason, the coarse-grained equilibrium bootstrap (3.32) is expected to perform well for the ground states while providing only modest improvements for the thermal states. It is plausible that a better coarse-grained bootstrap formulation specific to the thermal states based on MPO ansatz [4] exists, but we leave this investigation for future work.

## 4 Numerical results

We study the $(1+1)$-dimensional Transverse Field Ising Model (TFIM), whose Hamiltonian is given by
$$H_{\text{TFIM}} \equiv \sum_i H_i = -\sum_i Z_i Z_{i+1} - h \sum_i X_i, \tag{4.1}$$
where $X, Y, Z$ are Pauli matrices. The $h$ term is known as the transverse field term. We can also The system has a $\mathbb{Z}_2$ symmetry and is exactly solvable [44]. The model has a quantum critical point at $h = 1$, where the model is gapless, otherwise it is gapped.

- For $h < 1$ the $\mathbb{Z}_2$ symmetry is spontaneously broken and we get two degenerate vacua.

- For $h > 1$ the $\mathbb{Z}_2$ symmetry is preserved and the ground state is unique.

This model has energy denisty $\bar{E} \equiv \langle H_0 \rangle$ and magnetization per site $\langle Z_0 \rangle$ at zero-temperature given by [44]
$$\bar{E}_{\text{theory}} = -\frac{2(1+h)}{\pi} \, \text{E}\left(\frac{4h}{(1+h)^2}\right), \tag{4.2}$$

$$\langle Z_0 \rangle^{\text{theory}} = \begin{cases} (1-h^2)^{\frac{1}{8}} & \text{if } h \leq 1, \\ 0, & \text{if } h \geq 1, \end{cases} \tag{4.3}$$

where E is the elliptic integral of the second kind. $\langle Z_0 Z_1 \rangle$ correlator at arbitrary temperature $T = 1/\beta$ is given by [45]
$$\langle Z_0 Z_1 \rangle^{\text{theory}} = \frac{1}{\pi} \int_0^\pi \frac{1 + h \cos k}{\sqrt{1 + 2h \cos k + h^2}} \tanh\left(\beta \sqrt{1 + 2h \cos k + h^2}\right) dk. \tag{4.4}$$

A generalization of TFIM is a "detuned TFIM" where we explicitly break the $\mathbb{Z}_2$ symmetry by adding a longitudinal magnetic field with coupling $\lambda$
$$H_{\text{TFIM-}\lambda} \equiv H_{\text{TFIM}} - \lambda \sum_i Z_i, \tag{4.5}$$

If $\lambda$ is small, the phase diagram of detuned TFIM is related to the symmetric limit:

- When $h > 1, \lambda \neq 0$ only gives a perturbation to the order parameter.

- When $h < 1, \lambda \neq 0$ degeneracy of the vacua is lifted.

We also study XXZ model whose Hamiltonian is given by

$$H_{\text{XXZ}} \equiv \sum_i H_i = -\frac{1}{4}\sum_i \left(X_i X_{i+1} + Y_i Y_{i+1} + \Delta Z_i Z_{i+1}\right) \tag{4.6}$$

The XXZ system has a $U(1)$ symmetry. For a detailed review of the phase diagram of this model see eg. [46].

We use the setup of (3.32) to obtain the ground and thermal state properties of TFIM (4.1), such as the bound on energy density $\bar{E}$, order parameter $\langle Z_0 \rangle$ and $\langle Z_0 Z_1 \rangle$ correlator. We study the detuned TFIM (4.5) as an example of Hamiltonian without $\mathbb{Z}_2$ global symmetry. We also compute the ground state energy density of XXZ model (4.6). We see that the coarse-graining setup significantly narrows down the bootstrap bounds from [37] and in many cases, it helps obtain accurate prediction which would be exponentially difficult to obtain without coarse-graining. We also discuss the subtleties associated with $h < 1$ and the precision issue. Throughout the section, we use bond dimension $m = 2$. Since the dimension of the perturbative positivity matrix grows as $O(m^4)$, higher $m$ quickly becomes intractable in practice. We use numerical solver `mosek` [35] for most of the zero temperature problems and `sdpa-dd` [33, 47] for obtaining a few high-precision results.

We also provide preliminary results on the finite temperature setup of (3.32) for the TFIM Hamiltonian. The solver we use for the finite temperature case is `qics` [36].

### 4.1 Zero temperature bounds

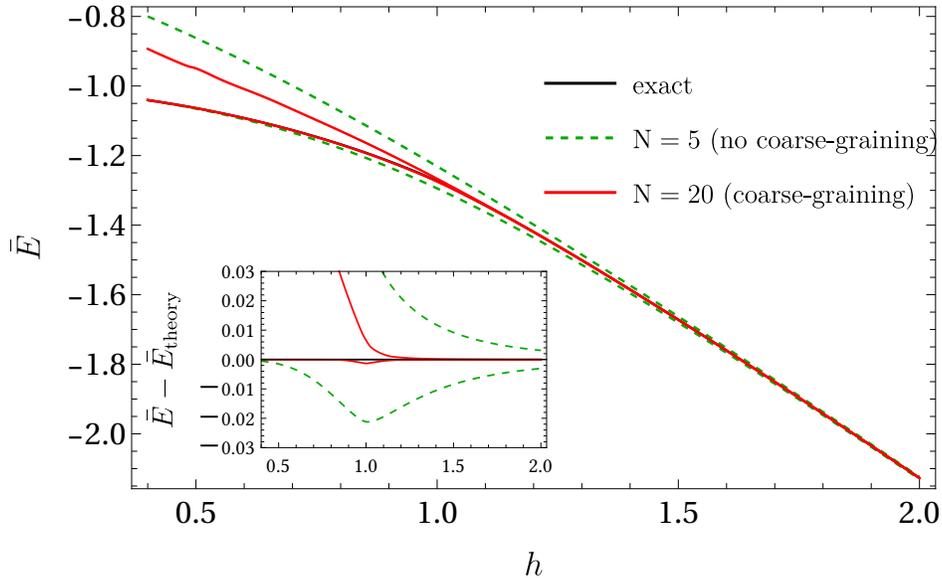

**Figure 1**. Upper and lower bounds on the energy density of TFIM (4.1) as functions of coupling $h$. The exact result $\bar{E}_{\text{theory}}$ is very close to the lower bounds.

First, we study the TFIM Hamiltonian (4.1) at zero temperature. Figure 1 shows bounds on the energy density $\bar{E}$ as a function of $h$. Previously, [37] computes both upper and lower bound at subsystem size $N = 5$ without using coarse-graining, and [38] computes the lower bound using coarse-grain and reaches large $N$. In our work, we are able to improve both upper bound and lower bound to $N = 20$. The new lower bound converges almost perfectly at all parameters $h$. The new upper bound converges perfectly in the symmetry preserving phase $h > 1$, and gets significantly closer than the previous bound in the broken phase $h < 1$.

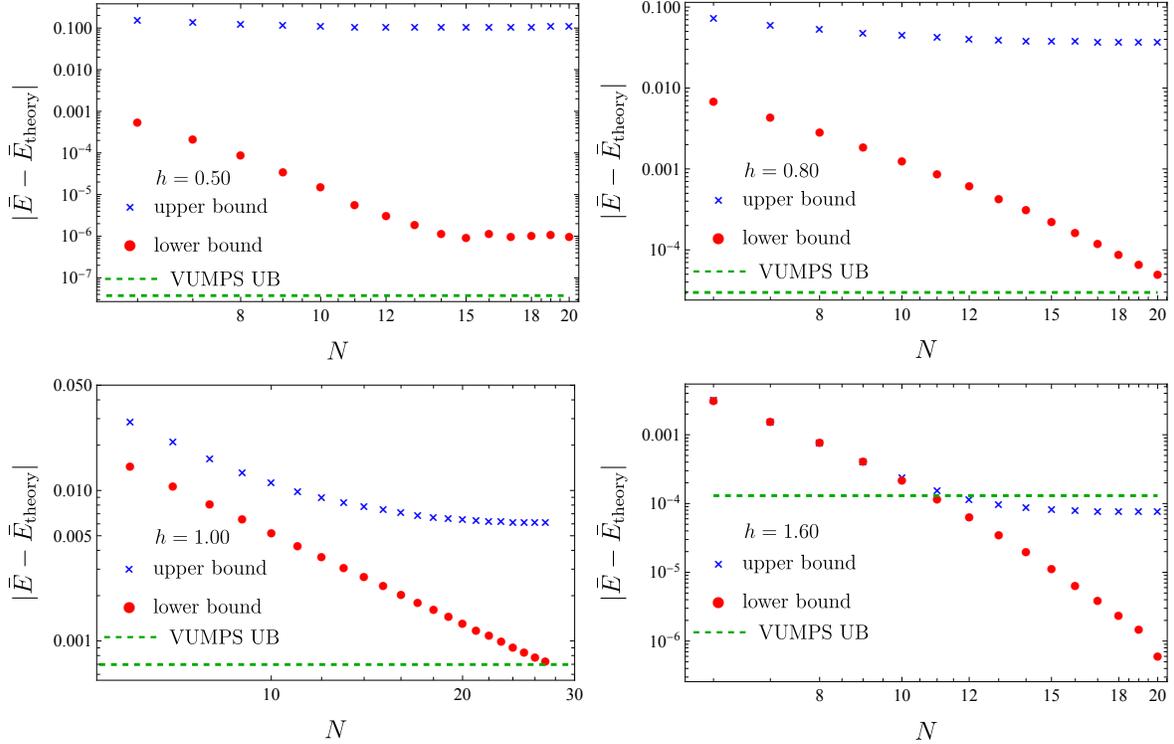

**Figure 2**. We show the difference between bounds of the energy density and exact values as the function of the number of sites in the sublattice $N$ for different values of $h$ in TFIM (4.1). We also show the upper bound (UB) obtained with the variational uniform MPS (VUMPS) algorithm [43] at bond dimension $m = 2$.

In Figure 2 we show the convergence of the new bounds as the subsystem size $N$ grows. The lower bound converges very fast to the exact ground state energy for non-critical coupling $h \neq 1$, all the way to $N = 20$. The $h = 0.5$ lower bound reaches a plateau at large $N$ likely because it reaches the numerical precision limit for double-precision solver `mosek`. At critical coupling, the convergence is slower and it agrees well with a power law. The upper bound goes down rapidly at small $N$, but slows down at larger $N$. For $h > 1$, the improved upper bound at $N = 20$ is orders of magnitude closer than the bound at $N = 5$. The bootstrap upper bound can be compared with the energy eigenvalue obtained using VUMPS method, which is also an upper bound on the ground state energy because VUMPS is a variational method.

In $h \gg 1$ parameter space, we see the end value of the bootstrap bound is comparable with the VUMPS method at the same bond dimension. For $h < 1$ the bound saturates at small $N$ and does not improve as much, indicating that the bootstrap setup encounters some difficulty when spontaneous symmetry breaking happens.

A key feature of our setup is that it provides high-precision two-sided bounds on any local observables. In Figure 3, we show the bounds on $\langle Z_0 \rangle$ and $\langle Z_0 Z_1 \rangle$ local operators. $\langle Z_0 \rangle$, aka the magnetization, is the order parameter of spontaneous $\mathbb{Z}_2$ symmetry breaking. The bound accurately predicts the magnetization at $h \gtrsim 1.3$ and $h \lesssim 1$. For $h \approx 1$ we are probing the near-critical behavior and the ability of the bootstrap setup to detect the critical point is limited by the correlation length. The coarse-grained setup allows us to efficiently bound the system with sublattice size up to $N \sim 20$, and the range is significantly narrowed down in the new result. The other operator, $\langle Z_0 Z_1 \rangle$, is a $\mathbb{Z}_2$ even operator. The behavior is similar to the energy density $\bar{E}$. The new bound at $N = 20$ significantly improves from the $N = 5$ previous results, and is almost saturated for the upper bound at $h \lesssim 1$ and both the upper and lower bound at $h \gtrsim 1.2$.

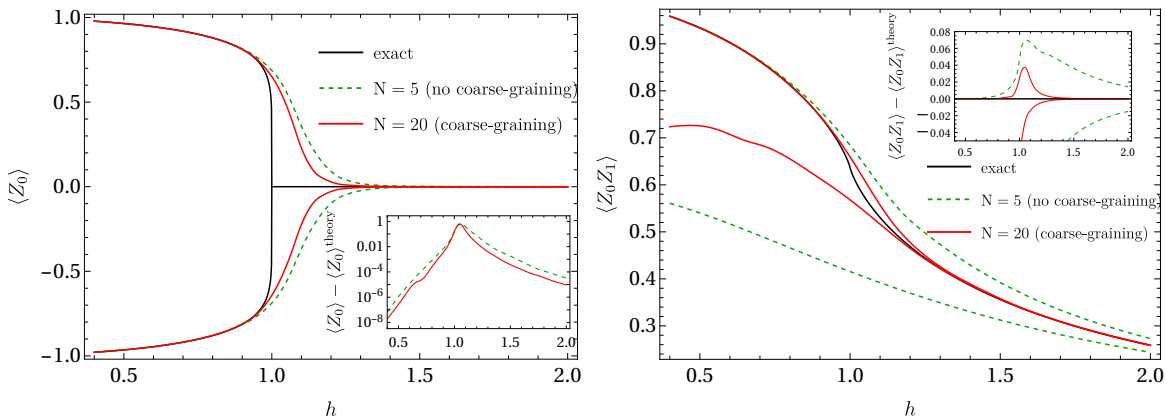

**Figure 3**. **Left Panel:** Upper and lower bounds on magnetization in TFIM (4.1) as functions of coupling $h$. **Right Panel:** Upper and lower bounds on correlator $\langle Z_0 Z_1 \rangle$ in TFIM as functions of coupling $h$.

The slow convergence in the broken phase $\lambda = 0, h < 1$ is puzzling. A possible explanation is that the perturbative positivity becomes inefficient when the ground state has degeneracy. This can be tested by turning on a small magnetic field $\lambda \neq 0$ in (4.1) to break $\mathbb{Z}_2$ symmetry and lift the degeneracy. Figure 4 shows that even a tiny magnetic field will immediately cause the upper bound to converge to the false vacuum energy. Now it seems that we answered one question but created another one: will the perturbative positivity eventually exclude the false vacuum? Schematically, the false vacuum is ruled out by the perturbative positivity

$$\langle \text{false vac} | c^\dagger [H, c] | \text{false vac} \rangle < 0 \tag{4.7}$$

if operator $c$ flips enough number of spins and destabilizes the false vacuum. For weak $\mathbb{Z}_2$

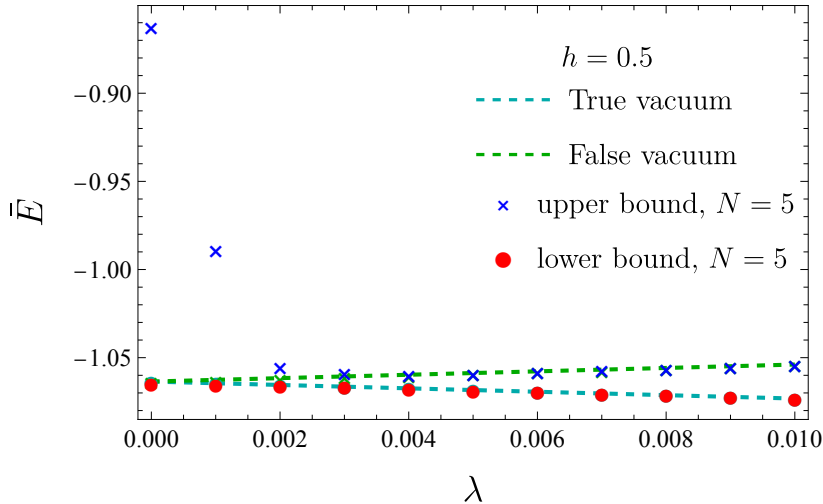

**Figure 4**. Bounds on the ground state energy density for TFIM (4.1) with a small magnetic field $\lambda \geq 0$. In the $\lambda = 0$ limit the ground state is degenerate and the upper bound is far from the ground state energy. As soon as we turn on a tiny magnetic field the upper bound converges very rapidly to the false vacuum energy, even for small subsystem size $N = 5$. The value of the true and false vacuum energies are approximated by the exact diagonalization result of the 16-site periodic spin chain.

breaking at small $\lambda$ it requires flipping many spins at once to see that the false vacuum is unstable, which puts a limit on the minimum $N$ to exclude the false vacuum. We test this picture in Figure 5, where we scan $N$ for a fixed transverse field $h$ and magnetic field $\lambda$. At $\lambda = 0.01$ bootstrap cannot exclude the false vacuum up to $N = 20$, whereas the $\lambda = 0.16$ energy upper bound has a sharp transition from the false vacuum energy to the true vacuum energy.

This result shows the power of coarse-graining. Without coarse-graining, [37], $N \sim 14$ bootstrap will be computationally infeasible even though the system carries a short correlation length. Coarse-graining, in contrast, takes full advantage of the short correlation length, pushing the bootstrap to higher $N$ and excluding the false vacuum.

In Figure 5 we face serious precision issues. Using double-precision solver `mosek` it is impossible to improve the bound much further than $N = 6$ and we would miss the entire transition behavior. We use `sdpa-dd`, a quadruple-precision solver, and obtained stable results which are consistent with even higher precision solver `sdpa-gmp`.

At a large magnetic field, the $\mathbb{Z}_2$ symmetry is strongly broken and the ground state is unique again. Table 1 shows that the bound rapidly converges to the vacuum energy in this regime.

Finally, we apply our setup on the XXZ model (4.6). The model has a $U(1)$ symmetry. The phase between $-1 \leq \Delta \leq 1$ is gapless and has a Berezinskii–Kosterlitz–Thouless phase transition at $\Delta = -1$. The phase at $\Delta < -1$ is anti-ferromagnetic. All these features indicate that the ground state properties may be significantly more difficult to bootstrap compared with TFIM. In Figure 6 we show our preliminary result for this model. The bound improves an $O(1)$ amount by going to $N \sim 10$ with a coarse-graining setup, but stalls for higher $N$. [38] shows that higher bond dimension $m > 2$ is required to obtain more precise lower bounds on ground state energy, and it is likely also true for upper bound on energy and bounds on other observables.

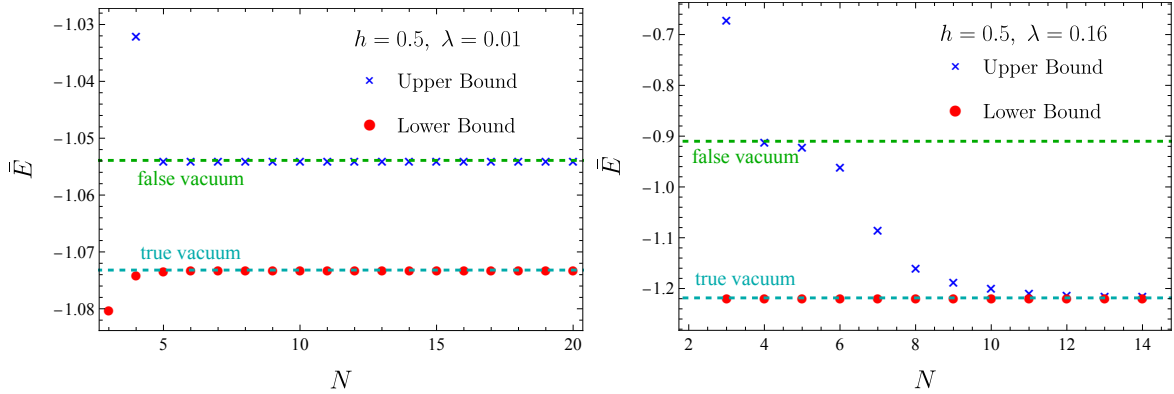

**Figure 5**. Upper and lower bounds on energy density in detuned TFIM (4.5) at $h = 0.5$, i.e. TFIM in the symmetry broken phase perturbed by a small magnetic field. The dashed lines show the approximate energy of the true ground state and false vacuum. The blue and red points show the upper and lower bound of the energy, respectively. The lower bound converges fast to the true vacuum. The upper bound first saturates at the false vacuum energy, but as $N$ grows it eventually converges to the true vacuum energy.

| $h$ | $N$ | lower bound $\langle H_0 \rangle$ | upper bound $\langle H_0 \rangle$ | VUMPS upper bound |
|---|---|---|---|---|
| 0.8 | 6 | -1.629001295 | -1.629001137 | |
| 0.8 | 7 | -1.629001171 | -1.629001135 | -1.6290011669134 |
| 1.5 | 6 | -1.958112676 | -1.958073470 | |
| 1.5 | 7 | -1.958084330 | -1.958075563 | |
| 1.5 | 8 | -1.958078303 | -1.958076062 | |
| 1.5 | 9 | -1.958076975 | -1.958076208 | -1.9580765814383 |

**Table 1**. We show the lower and upper bounds for the energy density in the detuned TFIM (4.5) with $\lambda = 0.5$. We also show the upper bound obtained with the VUMPS with bond dimensions up to $m = 8$. The precision of the obtained bounds is limited by the precision of the conic optimization solver.

**Numerical Performance** In Figure 7 we summarize the numerical performance of the semidefinite programming algorithm for the zero temperature energy problem defined in (3.32). The computation time scales with $N$ as a power law and fits to $\sim N^4$. We also compare with the run time without the perturbative positivity $\mathcal{C} \succeq 0$ constraints. If we remove $\mathcal{C} \succeq 0$, the problem essentially reduces to the problem in [38] which is targeted at the lower bound of the ground state energy. We see the bottleneck of the algorithm is the perturbative positivity, which is crucial for obtaining the upper bound on the ground state energy and both bounds for on any other local observables. The detailed definition of the semidefinite problem and numerical error are summarized in appendix B. Throughout the zero temperature section, we mostly use `mosek` with its default parameters, but for Figure 5 we use `sdpa-dd` for additional precision. In this case, we find that the primal feasibility error (B.3) threshold needs to be set lower (default $10^{-8}$, and we take $10^{-10}$). A double-precision solver stalls and cannot reach the error threshold for the given problem, and hence `sdpa-dd` is needed. The benchmark for these two solvers is summarized in Table 2.

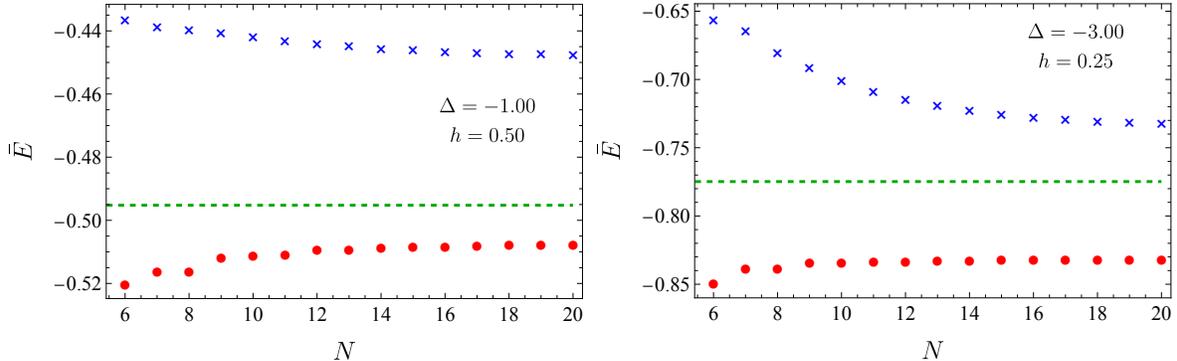

**Figure 6**. Upper and lower bounds on energy density in XXZ model as functions of subsystem size $N$. The dashed green line is the VUMPS upper bound computed with bond dimension $m = 2$.

|        | `mosek`    | `sdpa-dd`     |
|--------|------------|---------------|
| $N=6$  | 1.0 mins.  | 49.0 mins.    |
| $N=10$ | 27.3 mins. | 7.0 hours     |
| $N=14$ | 1.6 hours  | 1 day 4 hours |

**Table 2**. Benchmark for performance of energy upper bound problem in zero temperature (3.32) on `mosek` and `sdpa-dd`. While the former has sufficient precision in most of the parameter space, the latter is needed for the detuned TFIM Hamiltonian in the would-be-broken phase $h < 0$.

### 4.2 Finite temperature bounds

In this subsection, we discuss the numerical experiments of obtaining finite temperature bounds using the coarse-graining setup (3.32). As already discussed in section 3.4, thermal states are in general not well-approximated by MPS and rather require either MPO ansatz or a convex combination of MPS. Therefore, the coarse-grained equilibrium bootstrap (3.32) based on the uMPS, while providing a method to access bootstrap constraints at bigger subsystem sizes, does not guarantee bounds as tight as those for the ground states. Nonetheless, we present the coarse-grained bootstrap results for the thermal states here.

We take the Hamiltonian of TFIM (4.1) and try minimizing and maximizing the finite temperature expectation value $\langle Z_0 Z_1 \rangle_\beta$ for different coupling $h$ and inverse temperature $\beta$.

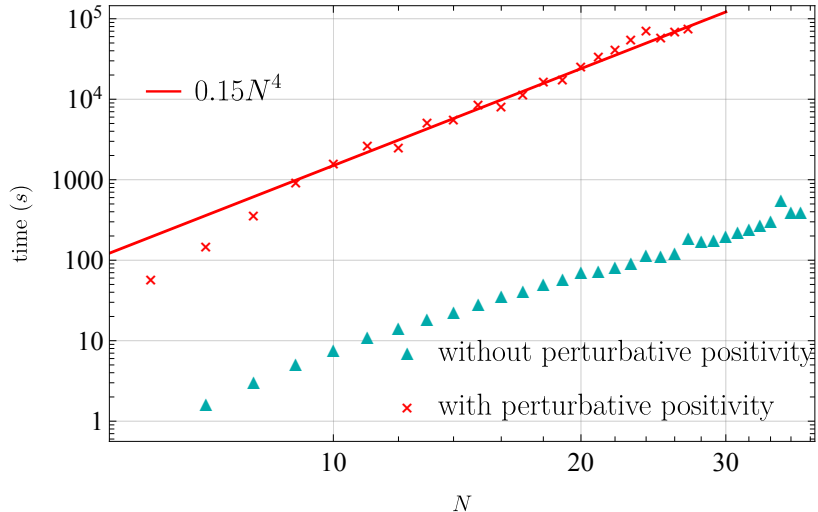

**Figure 7**. The run time of semidefinite programming with `mosek` for TFIM Hamiltonian (4.1) $h = 1$ without perturbative positivity (cyan) and with perturbative positivity (red). The former is computed on 2 Intel Xeon 6240 Processors, 18 cores per CPU, with 370 GB RAM. The latter is computed on 2 AMD EPYC 7742 CPUs, 64 cores per CPU, with 256 GB RAM. The performance depends very little on the number of CPUs.

We try both $N = 5$ with the no coarse-graining setup (3.31) and $N = 6$ with the coarse-graining setup (3.32). The former has been tested in [37] using Splitting Conic Solver. The latter is our main focus. For the coarse-graining setup, the uMPS that minimizes the zero temperature ground state energy is no longer expected to be optimal and in practice the uMPS associated with a slightly shifted $h$ may give a better bound. For each $h$, we try a few different uMPS at intermediate $\beta$, choose one uMPS based on the performance, and use it to scan all $\beta$. We use the solver `qics` [36] to efficiently compute the conic optimization problem using interior point method.

The result is shown in Figure 8. At high temperature $\beta \lesssim 1$, the bounds are close to the theoretical value. This is expected since the details of the Hamiltonian such as the correlation length does not affect the physics at high temperature due to the universal form of the density matrices. The only relevant length scale is provided by small $\beta$, and bootstrap constraints over a finite subsystem therefore produce very tight bounds. In this regime we see the coarse-graining bound at $N = 6$ has a significant improvement compared with $N = 5$.

At very low temperature $\beta \gtrsim 1$, the bounds are independent of $\beta$, because the density matrix is dominated by the ground state. We see the bound still has a sizable window at $N = 6$ but shrinks significantly compared with $N = 5$. The bound agrees with the zero temperature setup result as in the $\beta \to \infty$ limit the setup (3.32) reduces to the zero temperature setup. The intermediate temperature $\beta \sim 1$ converges the slowest. We see that coarse-graining helps improve the bound quantitatively. We do not know if the setup will keep converging at higher $N$. At $N = 5$ and $N = 6$, the conic optimization problem becomes quite expensive and requires an inverse Hessian-free setup provided by `qics` to have a manageable computation time and memory usage. On a Macbook Pro 2021, for each bound computed at $N = 6$ it takes about 2 hours for the solver to get close to a solution. Then, the program faces significant precision issues and struggles to reach a feasibility error $10^{-6}$. It is in many cases impossible to reach the default error $10^{-8}$.

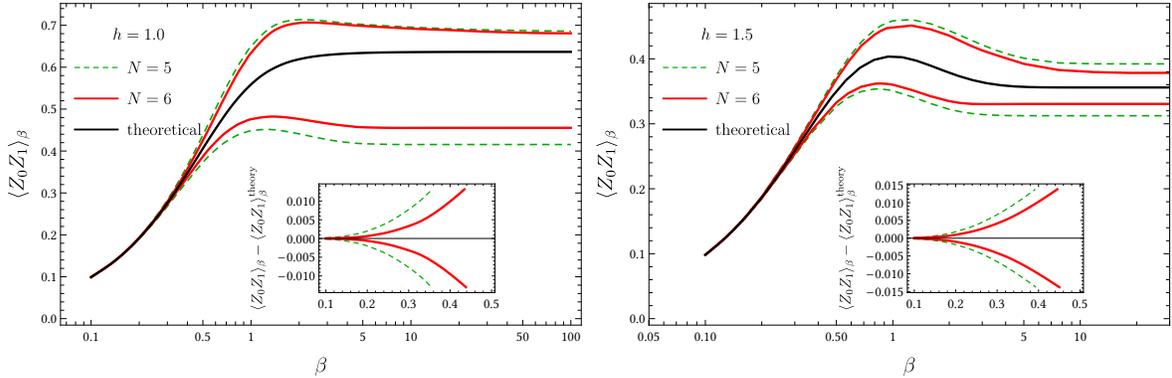

**Figure 8**. Bound on finite temperature expectation value $\langle Z_0 Z_1 \rangle_\beta$ in TFIM (4.1) at $h = 1.0$ and $h = 1.5$. For the former, we choose the uMPS of the ground state at $h = 1.0$ and for the latter, we use the uMPS of the ground state at $h = 1.4$. The red lines and blue lines show the bound at subsystem size $N = 5$ (using no coarse-graining setup) and $N = 6$ (using coarse-graining setup), respectively. The black line shows the theoretical value. At small $\beta$ the bounds are very close, and a zoomed-in plot of the bounds relative to the theoretical value is shown in the inset.

## 5 Discussion

For systems described by local Hamiltonians, it is generally expected that computing the relevant physics does not require access to the exponentially growing number of ingredients of the full system. Identifying systematic methods to extract the essential physics without accessing the complete configuration space is, therefore, crucial for achieving computational efficiency. A notable distinction between the ordinary bootstrap approach to many-body physics and other computational frameworks, such as tensor networks or Monte Carlo simulations, lies in their treatment of relevance: the latter inherently embed physical relevance in their formulations, while the former remains underdeveloped in strategies to isolate relevant ingredients. Resolving this disparity represents a conceptually significant challenge for the bootstrap philosophy.

In this work, we have made a progress by systematically incorporating the most relevant bootstrap constraints for quantum spin-chain systems using the uMPS coarse-graining approach. This allowed us to derive numerical upper and lower bounds on zero- and finite-temperature expectation values of local observables. The coarse-graining process enabled the inclusion of constraints from larger sublattices, yielding significantly tighter bounds compared to previous ones.

Even within the current coarse-grained equilibrium bootstrap formulation, natural next steps emerge. In the broken phase of TFIM, we observed less tight bounds compared to the preserving phase, likely due to the degeneracy of the ground states, which bootstrap constraints struggle to disentangle. However, introducing a longitudinal magnetic field to explicitly break this degeneracy immediately improved the bounds. We expect that incorporating global $\mathbb{Z}_2$ symmetry into the bootstrap framework, by decomposing bootstrap constraints according to irreducible representations, could address this issue.

Although our coarse-grained bootstrap framework utilized uMPS for thermal states, alternative bootstrap formulations, such as one based on the MPO ansatz, may provide a more tailored approach for thermal states. Such an exploration is left for the future.

Beyond the observables analyzed in this work, other key quantities in quantum many-body systems, such as the spectral gap, present compelling targets for future bootstrap analysis. Recent progress in bounding the spectral gap has been reported in [22, 48], but developing a systematic coarse-graining method for this problem could lead to significantly stronger bounds, as suggested in [48].

Finally, a fundamental question is whether the concept of coarse-graining can be applied to other types of bootstrap problems. Recent developments in lattice field theory bootstrap and quantum many-body bootstrap [10–14, 16, 27] face the common challenge of exponentially growing numbers of constraints. Identifying the most relevant subset of these constraints to capture the physics of interest remains an important and exciting direction for future investigation.

## Acknowledgements

We thank David Berenstein, Rajeev Erramilli, Hamza Fawzi, Omar Fawzi, Christian Ferko, Liam Fitzpatrick, Davide Gaiotto, Aleksandra Gočanin, Kerry He, Yin-Chen He, Ami Katz, Henry Lin, Zhijin Li, Maho Nakata, Joao Penedones, David Poland, Balt van-Rees, Slava Rychkov, Samuel O. Scalet, Grigory Tarnopolsky, Matthew Walters, Xi Yin, Bernardo Zan, Xiang Zhao and Yijian Zou for discussions. We are especially grateful to Hamza Fawzi for providing us with the data presented at [37]. The work of P.T. is supported by the Royal Society and by U.S. DOE grant DE-SC00-17660 and Simons Foundation grant #488651 (Simons Collaboration on the Nonperturbative Bootstrap). The work of Y.X. is supported by Simons Foundation grant #994316 for the Simons Collaboration on Confinement and QCD Strings and Yale Mossman Prize Fellowship in Physics. The work of Z.Z. is supported by Simons Foundation grant #994308 for the Simons Collaboration on Confinement and QCD Strings. We thank Bootstrap 2024 at the Complutense University of Madrid for the hospitality during the course of the work. C.N. and Y.X. thank the hospitality of Perimeter Institute during the preparation of the manuscript.

This work used the following computational facilities:

- Yale Grace computing cluster, supported by the facilities and staff of the Yale University Faculty of Sciences High Performance Computing Center.

- PSC bridges-2 cluster through allocation PHY240297 from the Advanced Cyberinfrastructure Coordination Ecosystem: Services & Support (ACCESS) program, which is supported by U.S. National Science Foundation grants #2138259, #2138286, #2138307, #2137603, and #2138296.

## A  Energy-entropy balance inequalities

In this appendix, we justify the following Energy-Entropy Balance inequalities for the thermal ensemble $\rho = \exp(-\beta H)$:

$$\beta \langle \mathcal{O}^\dagger [H, \mathcal{O}] \rangle_\beta \geq \langle \mathcal{O}^\dagger \mathcal{O} \rangle_\beta \log \left( \frac{\langle \mathcal{O}^\dagger \mathcal{O} \rangle_\beta}{\langle \mathcal{O} \mathcal{O}^\dagger \rangle_\beta} \right) \tag{A.1}$$

for any operator $\mathcal{O}$, following the ideas presented in [37, 41]. This condition, combined with additional conditions, can not only be necessary but also sufficient if it holds for any operator $\mathcal{O}$, but in this work, we only rely on the necessity. Here, in this appendix, the expectation $\langle A \rangle_\beta$ refers to the expectation under the density matrix $\mathrm{Tr}(\rho A)$.

The general idea is to utilize the convexity of the right-hand side of inequality Eq. (A.1):

$$S(u, v) = u \log(u/v) \tag{A.2}$$

To see that this function is convex, we compute the Hessian matrix of $S(u, v)$:

$$\begin{pmatrix} \frac{1}{u} & -\frac{1}{v} \\ -\frac{1}{v} & \frac{u}{v^2} \end{pmatrix} \succeq 0, \quad \text{for } u, v > 0, \tag{A.3}$$

Moreover, the function is homogeneous in $(u, v)$:

$$S(\lambda u, \lambda v) = \lambda S(u, v) \tag{A.4}$$

Thus, if we could construct a positive measure $\mathrm{d}\mu$ such that:

$$\int \mathrm{d}\mu f_1 = \langle \mathcal{O}^\dagger \mathcal{O} \rangle_\beta, \quad \int \mathrm{d}\mu f_2 = \langle \mathcal{O} \mathcal{O}^\dagger \rangle_\beta, \quad \int \mathrm{d}\mu S(f_1, f_2) = \beta \langle \mathcal{O}^\dagger [H, \mathcal{O}] \rangle_\beta \tag{A.5}$$

Then, by the convexity property, we have:

$$\int \mathrm{d}\mu S(f_1, f_2) \geq S\left( \int \mathrm{d}\mu f_1, \int \mathrm{d}\mu f_2 \right) \tag{A.6}$$

which proves the positivity condition (A.1).

To achieve this, we introduce the following auxiliary Hilbert space, where the vectors are operators of the quantum system under investigation, and the inner products are defined by:

$$(\mathcal{O}_2, \mathcal{O}_1) = \langle \mathcal{O}_2^\dagger \mathcal{O}_1 \rangle_\beta \tag{A.7}$$

In this Hilbert space, the operator $L$ defined by $L\mathcal{O} = [H, \mathcal{O}]$ is self-adjoint:

$$(\mathcal{O}_2, L\mathcal{O}_1) = \langle \mathcal{O}_2^\dagger [H, \mathcal{O}_1] \rangle_\beta = \langle [\mathcal{O}_2^\dagger, H] \mathcal{O}_1 \rangle_\beta = (L\mathcal{O}_2, \mathcal{O}_1) \tag{A.8}$$

Here, we used the condition that $\langle H\mathcal{O} \rangle_\beta = \langle \mathcal{O}H \rangle_\beta$. We can now expand $L$ using spectral decomposition:

$$L = \sum_i \lambda_i |\psi_i)(\psi_i|, \quad \lambda_i \in \mathbb{R} \tag{A.9}$$

where $|\psi_i)(\psi_i|$ is the projection operator in the auxiliary Hilbert space, satisfying the completeness condition:

$$\mathbb{I} = \sum_i |\psi_i)(\psi_i| \tag{A.10}$$

and $L|\psi_i) = \lambda_i |\psi_i)$.

Define $\mu_i = |(\psi_i, \mathcal{O})|^2$, then we have:

$$\langle \mathcal{O}^\dagger \mathcal{O} \rangle_\beta = \sum_i \mu_i, \tag{A.11}$$

$$\beta \langle \mathcal{O}^\dagger [H, \mathcal{O}] \rangle_\beta = \beta(\mathcal{O}, L\mathcal{O}) = \beta \sum_i \mu_i \lambda_i = \sum_i \mu_i S(1, \exp(-\beta \lambda_i)) \tag{A.12}$$

$$\langle \mathcal{O}\mathcal{O}^\dagger \rangle_\beta = (\mathcal{O}, \exp(-\beta L)\mathcal{O}) = \sum_i \mu_i \exp(-\beta \lambda_i), \tag{A.13}$$

The first equal sign in the above formula (A.13) comes from:

$$\langle \mathcal{O}_1 \mathcal{O}_2 \rangle_\beta = \text{Tr}(\rho \mathcal{O}_1 \mathcal{O}_2) = \text{Tr}(\mathcal{O}_2 \rho \mathcal{O}_1) = \text{Tr}\left(\rho \mathcal{O}_2 e^{-\beta H} \mathcal{O}_1 e^{\beta H}\right) = \langle \mathcal{O}_2 e^{-\beta H} \mathcal{O}_1 e^{\beta H} \rangle_\beta \tag{A.14}$$

This condition is usually called the KMS condition in the literature [41].

The above three equations (A.11), (A.12) and (A.13), combined with the convexity condition (A.6), finally prove the Energy-Entropy Balance inequalities (A.1).

## B  The Primal-Dual Interior-Point Algorithm of Semidefinite Programming

In this appendix, we review some basics of the implementation of the primal-dual interior-point algorithm to solve semidefinite programming (SDP). Our conventions are based on sdpa [49] and sdpb [34]. For more general implementations of the SDP problem, readers are encouraged to refer to [50, 51].

In our convention, the primal and dual form of the SDP is defined as:

$$\left.\begin{array}{l} \mathcal{P}\text{: minimize } \sum_{i=1}^m c_i x_i \text{ subject to } \boldsymbol{X} = \sum_{i=1}^m \boldsymbol{F}_i x_i - \boldsymbol{F}_0, \quad \boldsymbol{X} \succeq \boldsymbol{O}, \quad \boldsymbol{X} \in \mathbb{S}^n. \\ \mathcal{D}\text{: maximize } \boldsymbol{F}_0 \bullet \boldsymbol{Y} \quad \text{subject to } \boldsymbol{F}_i \bullet \boldsymbol{Y} = c_i \quad (i = 1, 2, \ldots, m), \quad \boldsymbol{Y} \succeq \boldsymbol{O}, \quad \boldsymbol{Y} \in \mathbb{S}^n. \end{array}\right\} \tag{B.1}$$

The primal and dual interior-point algorithm is based on the strong duality of the primal and dual problems (Slater's theorem), which is guaranteed if either the primal or dual problem is strictly feasible (the interior of the feasible set is non-empty). The interior-point algorithm for SDP is implemented by applying Newton's method to solve the following KKT system:

$$\boldsymbol{X} = \sum_{i=1}^{m} \boldsymbol{F}_i x_i - \boldsymbol{F}_0, \quad \boldsymbol{F}_i \bullet \boldsymbol{Y} = c_i \quad (i=1,2,\ldots,m), \quad \boldsymbol{XY} = \boldsymbol{O}, \quad \boldsymbol{X} \succeq \boldsymbol{O}, \quad \boldsymbol{Y} \succeq \boldsymbol{O} \quad \text{(B.2)}$$

In practice, we begin with $\boldsymbol{X}$ and $\boldsymbol{Y}$ as positive semidefinite matrices and then use Newton's method to satisfy the equality conditions in Eq. (B.2) while maintaining positive semidefiniteness. The iterations end when both the primal/dual feasibility and the duality gap are below a given threshold, which is set as an input parameter.

In SDPA [49], the primal and dual feasibility are defined as:

$$\max\left\{\left|\left[\boldsymbol{X} - \sum_{i=1}^{m} \boldsymbol{F}_i x_i + \boldsymbol{F}_0\right]_{pq}\right| : p,q = 1,2,\ldots,n\right\}, \quad \text{(B.3)}$$

$$\max\left\{|\boldsymbol{F}_i \bullet \boldsymbol{Y} - c_i| : i = 1,2,\ldots,m\right\},$$

respectively, while the duality gap is defined as:

$$\frac{|\text{objP} - \text{objD}|}{\max\left\{\frac{(|\text{objP}|+|\text{objD}|)}{2.0}, 1.0\right\}}, \quad \text{(B.4)}$$

where "objP" and "objD" denote the objective values of the primal and dual problems, respectively.

## C  More convergence plots

In this appendix, we show more convergence plots. Figure 9 shows the convergence of the energy density in the detuned TFIM (4.5) for $h = 1.5$ and different values of longitudinal field coupling $\lambda$. For $N = 20$, our upper bound is below the VUMPS upper bound.

Figure 10 shows the difference between bounds on magnetization per site $\langle Z_0 \rangle$ and the theoretical value for TFIM (4.1), in broken phase $h = 0.8$ and in the preserving phase $h = 1.6$. Non-monotonicity of the bounds as functions of $N$ is the consequence of the numerical fluctuations of the solver.

Figure 11 shows the difference between bounds and theoretical value of correlator $\langle Z_0 Z_1 \rangle$ as the function of sublattice size $N$ in TFIM (4.1). In the preserving phase, bounds converge faster to the exact value than in the broken phase.

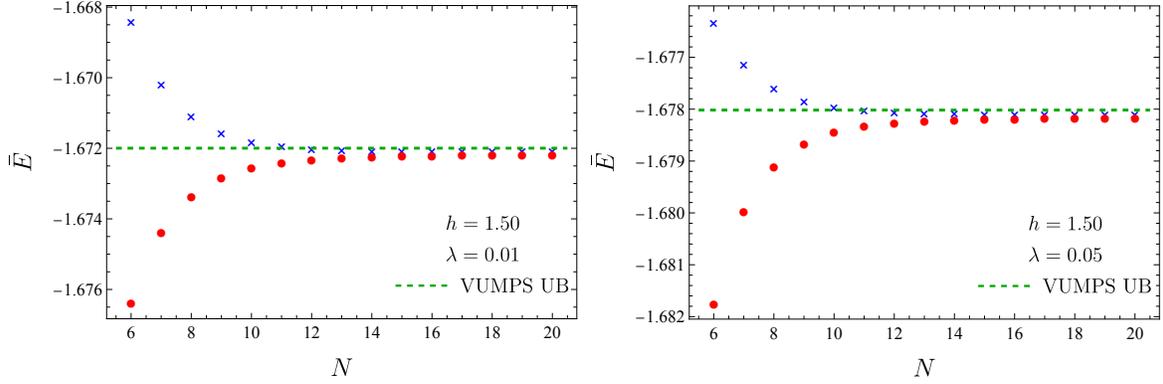

**Figure 9.** Upper and lower bounds on energy density in detuned TFIM (4.5) as functions of subsystem size $N$. The dashed green line is the VUMPS upper bound computed with bond dimension $m = 2$.

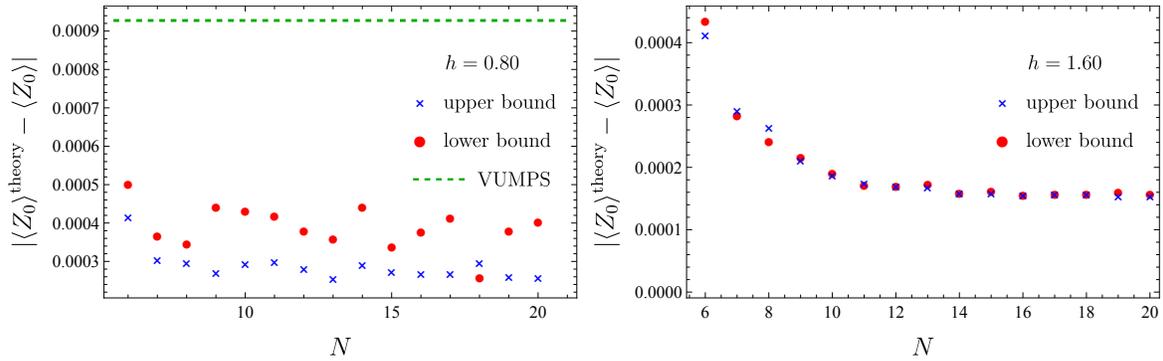

**Figure 10.** We show the difference between bounds on magnetization per site $\langle Z_0 \rangle$ and the exact values as the function of the sublattice size $N$ in TFIM (4.1). We also show the values obtained with the VUMPS [43] algorithm in the broken phase state. In the phase-preserving state, the VUMPS value matches the exact value up to the numerical fluctuations. Non-monotonicity of the bounds as functions of $N$ is the consequence of the numerical fluctuation of the solver.

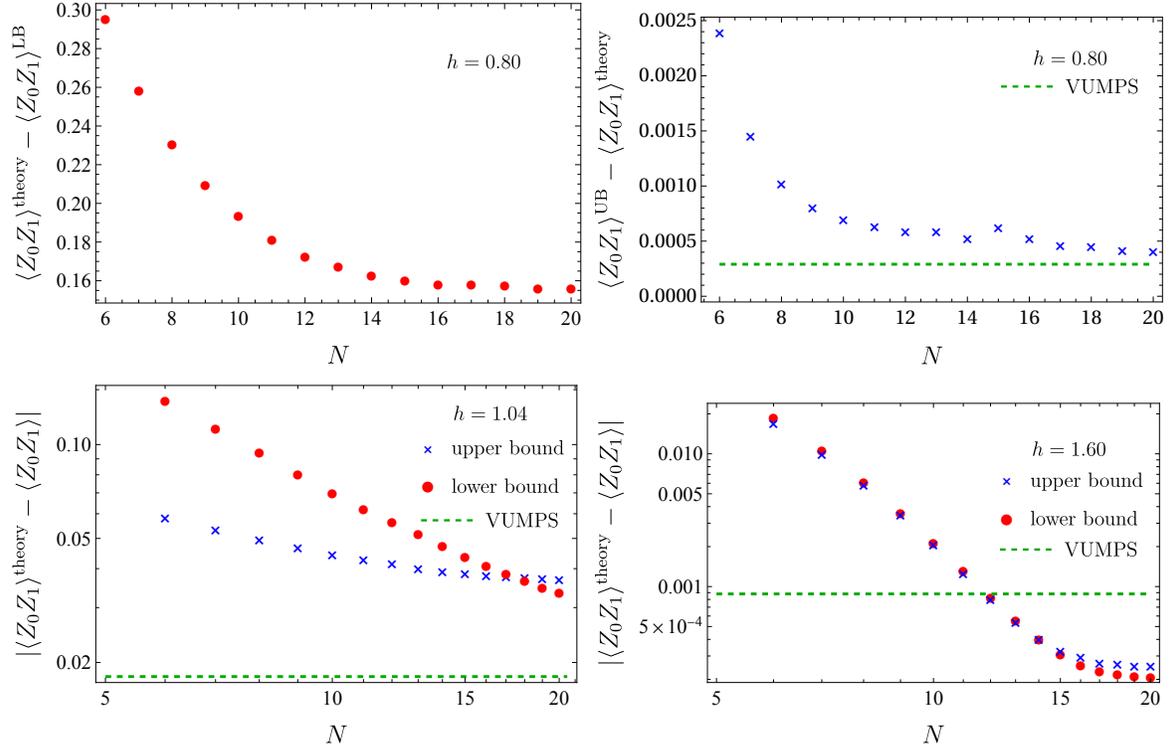

**Figure 11**. We show the difference between bounds on $\langle Z_0 Z_1 \rangle$ correlator and the exact result as the function of the sublattice size $N$ in TFIM (4.1). We also show the values obtained with the VUMPS. Our bounds are more accurate than the VUMPS result in the phase-preserving state.